\documentclass[pra,aps,twocolumn,superscriptaddress,nofootinbib,longbibliography]{revtex4-2}
\usepackage{graphicx}
\usepackage{dcolumn}
\usepackage{bm}
\usepackage{amsmath}
\usepackage{epsfig}
\usepackage{color}
\usepackage{braket}

  \def\kb{{\rm \bf k}} \def\pb{{\rm \bf \hat{p}}} \def\rb{{\rm \bf r}}   

  \def\qb{{\rm \bf q}}  \def\Ab{{\rm \bf A}} \def\Bb{{\rm \bf B}}

 \def\Eb{\vec{\mathcal{E}}}

\begin{document}
\title{Spin flips of free electrons in optical near fields}
\author{Deng~Pan}
\email[Corresponding author: ]{dpan@lps.ecnu.edu.cn}
\affiliation{State Key Laboratory of Precision Spectroscopy, East China Normal University, Shanghai 200062, China}
\author{Hongxing~Xu}
\email[Corresponding author: ]{hxxu@whu.edu.cn}
\affiliation{School of Physics and Technology, Wuhan University, Wuhan 430072, China}


\begin{abstract}
Manipulating the spin polarization of electron beams using light is highly desirable but exceedingly challenging, as the approaches proposed in previous studies using free-space light usually require enormous laser intensities. Here, we propose the use of a transverse electric optical near field, extended on nanostructures, to efficiently induce spin flips of an adjacent electron beam by exploiting the strong inelastic electron scattering in phase-matched optical near fields. Our calculations show that the use of a dramatically reduced laser intensity ($\sim 10^{12}\,$W/cm$^2$) with a short interaction length ($16\,\mu$m) achieves an electron spin-flip probability of approximately $12\%$. Intriguingly, the two spin components of an unpolarized incident electron beam---parallel and antiparallel to the electric field---are spin-flipped and inelastically scattered to different energy states, providing an analog of the Stern--Gerlach experiment in the energy dimension. Our findings are important for optical control of free-electron spins, preparation of spin-polarized electron beams, and applications as varied as in material science and high-energy physics. 
\end{abstract}
\date{\today}

\maketitle
 
\section{Introduction}
After Stern and Gerlach's celebrated experiment using inhomogeneous magnetic fields to separate neutral atoms by spin, Bohr and Pauli pointed out that a similar scheme could not be simply applied to polarize charged particles due to the influence of the Lorentz force \cite{P1932}. Consequently, alternative mechanisms are required to prepare spin-polarized electrons \cite{K1985}. Two main approaches have been implemented experimentally: one uses semiconductor photocathodes with negative electron affinities as electron sources \cite{PMZ1975,PM1976,PCW1980}, while the other is based on spin-dependent radiation from relativistic electron beams in storage rings (i.e., the Sokolov--Ternov effect) \cite{ST1964,ST1967,B99}. Importantly, spin-polarized electron beams have served as an essential tool for investigating the magnetic properties of solid-state materials and molecules \cite{F1986_3, K08_3,G10_2}, probing atom and nucleon spin structures \cite{J97,G09_2, ABS13}, and studying fundamental problems in high-energy physics \cite{PAC1978,AAA04,AAA16}.

Despite a few works still examining the possibility of separating different electron spins by static magnetic fields \cite{BGS97, RGG98, RG98, GBG01,MBB11}, most concentrated efforts are now being devoted to exploring the possibility of polarizing electrons using free-space light. In these latter studies, the spin dynamics of electrons are manipulated via Compton scattering \cite{WUH02, DSB17, SDR18,LSH19,SWL19,LCH22} or by the Kapitza--Dirac effect \cite{FB03,R04,ABK12,ABK13,MHS15,DAM16,DM17,A17,EVS18,ALC20,AGS22}, which involve either a single laser beam or two counter-propagating lasers \cite{KD1933,B07,FAB01}, respectively.
Nevertheless, these free-space phenomena are at least second-order quantum processes, which are significant only at enormous laser intensities (typically, $\sim 10^{18}$--$10^{22}\,$W/cm$^2$, depending on the interaction time; see Appendix). More efficient first-order quantum processes are simply forbidden by the energy-momentum conservation laws. More precisely, in free space, it is not possible to switch the spin AM (angular momentum) of an electron between $-\hbar/2$ and $\hbar/2$ by directly absorbing or emitting a photon with an AM of $\hbar$. 

In contrast to these free-space scenarios, interactions between moving electrons and optical modes in media or localized structures are intrinsically inelastic, mediated by large photonic momenta in the near field. For example, in Cherenkov radiation (CR) and electron energy-loss spectroscopy (EELS), electron beams spontaneously release photons into vacuum photonic modes. Recently, the photon-induced near-field electron microscopy (PINEM) technique has been intensively investigated \cite{BFZ09,paper151,PLZ10,PZ12,KGK14,PLQ15,FES15,paper282,EFS16,VFZ16,KML17,FBR17,PRY17,RB16,PRY17,MB18,paper311,DNS20,KLS20,DGH21,SCH22}. Unlike those spontaneous-emission-like phenomena such as CR and EELS, a laser is introduced in PINEM to excite the optical near field of a nanostructure. At moderate laser intensity, the probability of an electron being inelastically scattered while moving through the optical near field can reach unity. Surprisingly, by tailoring the sample geometry and illumination to match the electron velocity and near-field phase velocity, recent PINEM experiments have observed emissions or absorptions of hundreds of photons by a single electron \cite{DNS20,KLS20}. Given this strong inelastic scattering of electrons experienced in PINEM, it is thus illuminating to consider whether electron spin-flip transitions can be efficiently achieved by direct AM exchanges with optical near fields.

In this work, we reveal a pathway for achieving significant spin flips of electron beams via interactions with a transverse electric (TE) optical near field by exploiting the efficient inelastic electron scattering in the optical near field as in PINEM. Compared to previous studies that involves only free-space light, the intensity of the laser required to induce pronounced electron spin-flip transitions in our approach is dramatically reduced ($\sim 10^{12}\,$W/cm$^2$ in our calculations). As the interactions between the electron beam and optical near field are intrinsically inelastic, the electron spin can be flipped by directly absorbing or emitting a photon of momentum $\hbar$. Our calculations demonstrate that the desired electron spin-flip transitions can be achieved along an arbitrary direction of concern. More importantly, for an unpolarized incident electron beam, the two spin components along the electric field direction are scattered to different energy states, thereby offering an optical way to prepare spin-polarized electron beams.

\begin{figure}[t]
	\begin{centering}
		\includegraphics[width=0.5\textwidth]{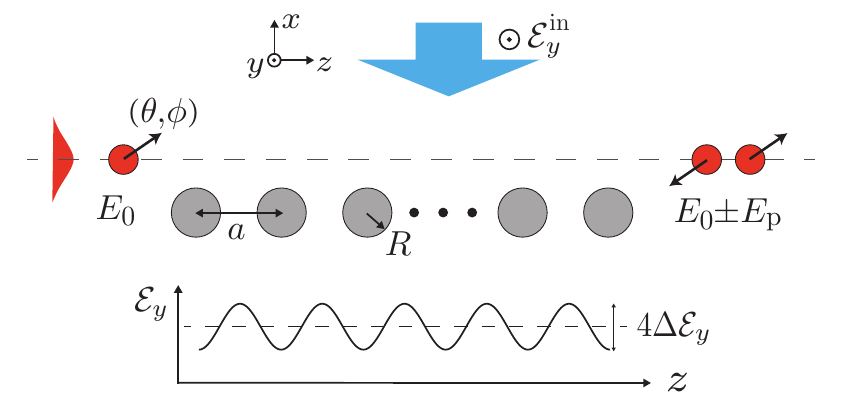}
		\par\end{centering}
	\caption{ Illustration of electron spin flips in an optical near field. A wide laser beam (blue arrow) is normally incident on a nanowire array, with the electric field parallel to the nanowires. This results in a periodically modulated near field (bottom panel). An electron passing through the optical near field is scattered by absorbing or emitting a photon of energy $E_{\rm p}$, while its spin can be flipped.
 }
	\label{Fig1}
\end{figure}

\section{Model system}
In this work, we consider a periodic two-dimensional (2D) nanostructure normally illuminated by a laser, as exemplified by a nanowire array (see Fig.\ \ref{Fig1}). The excited optical near field can induce inelastic and spin-flip scattering of an adjacent electron beam (propagating along $+\hat{z}$). For simplicity, we also assume a 2D electron beam, with its y-component wave vector vanishing. In PINEM, the electric field component parallel to the electron beam dominates, which induces only spin-preserving transitions. In the present work, we consider a TE configuration (only $\mathcal{E}_y$ is present, see Fig.\ \ref{Fig1}) in which the spin-flip transitions can occur (see below). Regarding the nanowire array under such TE illumination, the total electric field includes a near-field component and can be approximated as $\vec{\mathcal{E}}_y(\rb)= \vec{\mathcal{E}}_y^{\rm in}(x) +2\Delta\vec{\mathcal{E}}_y (x)\cos(gz)$, where $g=2\pi/a$. We also assume a monochromatic light field for analysis, i.e., $\vec{\mathcal{E}}_y(\rb,t)=\vec{\mathcal{E}}_y(\rb)e^{-i\omega t}+c.c.$, which can be easily extended to a pulsed laser field, as in the PINEM theory \cite{DNS20}.

As in a PINEM  experiment \cite{DNS20,KLS20},  we  guarantee  a strong electron--photon interaction by considering a resonant situation throughout this work, assuming that the initial electron velocity matches the near-field phase velocity, i.e., $\beta= v_0/c=a/\lambda$, where $\lambda$ is the vacuum photon wavelength. In contrast to the high-energy electron beams ($>100\,$keV) usually adopted in PINEM (see exceptions in \cite{T20,SCH22}), as we shall explain below, the spin-flip transitions of the electrons are only significant for low-energy electrons. We also note that such a low-energy electron beam is not as tightly bounded as in PINEM (see red shade in Fig.\ \ref{Fig1}). This wide electron beam has an additional complication that the high laser intensity required to achieve significant spin-flip transitions can also cause an electron-beam reshaping effect. Our theory, presented below, can fully capture both the spin dynamics and diffraction effects of electron beams in the optical near field.

\section{Theory of spin dynamics}
As the spin-flip effect is significant only for low electron velocities (see below), we restrict this work to nonrelativistic situations. Given this limit, the spin dynamics of an electron beam is governed by Pauli’s equation, $(\pb^2/2 m_e+\mathcal{\hat{H}}_{\rm I})\Psi(\rb,t)=i\hbar\partial_t\Psi(\rb,t)$ ($\rb=(x,z)$), where the interaction Hamiltonian reads 
\begin{align}
	\mathcal{\hat{H}}_{\rm I}=\frac{e}{m_e c}\Ab(\rb,t)\cdot\pb + \mu_B \Bb(\rb,t)\cdot\vec{\sigma} +\frac{e^2}{2m_e c^2} \Ab^2(\rb,t) ,\nonumber
\end{align} 
$\pb$ is the momentum operator, $\mu_B= e\hbar/2 m_e c$ is the Bohr magneton, and $\vec{\sigma}=[\sigma_x,\sigma_y,\sigma_z]$ is the three-vector Pauli matrix. The wave function in Pauli's equation is described using the two-component spinors, $\Psi(\rb,t)=\sum_{\pm} \Psi^\pm(\rb,t)\otimes\ket{\pm} $, where $\ket{+}=[\cos(\theta/2),\sin(\theta/2)e^{i\phi}]^T$ and $\ket{-}=[\sin(\theta/2),-\cos(\theta/2)e^{i\phi}]^T$ denote the two possible spins of electrons along the direction implicitly defined by the two polar coordinates $(\theta,\phi)$ (see Fig.\ \ref{Fig1}). To derive the interaction Hamiltonian $\mathcal{\hat{H}}_{\rm I}$ above, a radiation gauge is chosen so that the scalar potential vanishes and the vector potential is related to the electric field by $\Ab(\rb)=-i\mathcal{E}(\rb)/q$, where $q=\omega/c$. In what follows, we refer to the three terms in $\mathcal{\hat{H}}_{\rm I}$, from left to right, as $\mathcal{\hat{H}}_{\rm I}^{(1)}$, $\mathcal{\hat{H}}_{\rm I}^{(2)}$, and  $\mathcal{\hat{H}}_{\rm I}^{(3)}$, respectively. 

Considering an electron beam of initial energy $E_0$ interacting with an optical near field, the electron energy should spread only among discrete levels, i.e., $E_n=E_0+ n \hbar\omega$. In the case of a paraxial and slowly varying electron beam, we can expand the electron wave function as $\Psi^\pm (\rb,t)= \sum_n \psi_n^\pm (\rb) e^{i(p_n z-E_n t)/\hbar}$, where $p_n=\sqrt{2m_e E_n} \approx p_0 + n\hbar \omega/v_0$. Using this expansion, and together with a nonrecoil approximation, Pauli's equation can be rewritten as a set of coupled equations (see Appendix),
\begin{align}
	\big( v_0 \hat{p}_z  +\frac{\hat{p}^2_x}{2 m_e}\big) \psi_n^s(\rb)= - \sum_{n's'} \mathcal{M}^{ss'}_{nn'} \psi_{n'}^{s'}(\rb) ,\label{Schr2}
\end{align} 
where the nonzero transition matrix elements for the TE electromagnetic field are
\begin{align}
	\mathcal{M}_{n, n-1}^{ss'}&= e^{-i \frac{\omega}{v_0} z} \mu_B \bra{s} \Bb(\rb) \cdot\vec{\sigma})\ket{s'}  , \label{offDmat} \\ 
    \mathcal{M}_{n n}^{ss}&= \frac{e^2 }{ m_e \omega^2} |\mathcal{E}_y(\rb)|^2      ,\nonumber
\end{align} 
and their Hermitian conjugates $\mathcal{M}_{n'n}^{s's}=\big(\mathcal{M}_{nn'}^{ss'}\big)^*$. The above matrix elements reveal that the magnetic field can induce both spin-flip and spin-preserving transitions. In addition, the interaction term $\mathcal{\hat{H}}_{\rm I}^{(3)}$, which corresponds to the ponder-motive force, introduces an additional phase to the wave function \cite{paper368}.

As seen from Eq.\ (\ref{offDmat}), for the infinitely extended nanostructure shown in Fig.\ \ref{Fig1}, resonantly enhanced interactions arise when $\omega/v_0=g$ or equivalently $v_0/c=a/\lambda$ \cite{DNS20,KLS20}. Additionally, in Eq.\ (\ref{offDmat}), only the Fourier component of the near field with momentum along the z-direction equal to $g$, e.g., the electric field component $\Delta\vec{\mathcal{E}}_y(x) e^{i g z} $ (see Fig.\ \ref{Fig1}), contributes to the electron dynamics. Such Fourier component is an evanescent wave characterized by a factor $e^{-q_x x}$, where $q_x=\pm\sqrt{g^2-q^2}$. We assume that the electron beam is above the nanowires, i.e., $q_x > 0$ (see Fig.\ \ref{Fig1}), so the corresponding magnetic field component is $\Delta \Bb(\rb)=\qb\times\Delta\vec{\mathcal{E}}_y(x) e^{ig z}/ q$, according to Maxwell's equations, where $\qb=(i q_x,0, g)$. Using the magnetic field component $\Delta \Bb(\rb)$, we can further simplify the off-diagonal matrix elements in Eq.\ (\ref{offDmat}) to
\begin{align}
	\mathcal{M}_{n, n+1}^{s's}= \mu_B [ \bra{s}(\qb\times \hat{y})\cdot\vec{\sigma}\ket{s'}]\Delta\mathcal{E}_y(x)/q  .\label{Mmat}
\end{align} 
Importantly, we note that $\Delta \Bb(\rb)$ is circularly polarized when $g\gg q$, which is associated with the spin AM in the evanescent near field \cite{BBN14_2,BSN15}. Meanwhile, when $g\gg q$, the above matrix element is approximated by $\mathcal{M}_{n, n+1}^{s's}\approx 2\mu_B \Delta \mathcal{E}_y \beta^{-1}$ for spin flips along the y-direction.

In the weak-interaction regime, the spin-flip transition probabilities over a propagation distance of $L$ can be evaluated, according to Eqs.\ (\ref{Schr2}) and (\ref{Mmat}), by $|2\mu_B  \Delta \mathcal{E}_y L/c\hbar|^2 \beta^{-4}$, which is greatly increased when electrons slow down. However, an excessively small electron velocity widens the electron beam waist and requires a more localized near field ($\sim e^{-g x}$) to fulfill the phase-match condition, thereby reducing the chance of electron--photon interaction. Consequently, we assume a moderate electron velocity, i.e., $\beta=1/10$, in the detailed calculations below. 

\begin{figure}[t]
	\begin{centering}
		\includegraphics[width=0.5\textwidth]{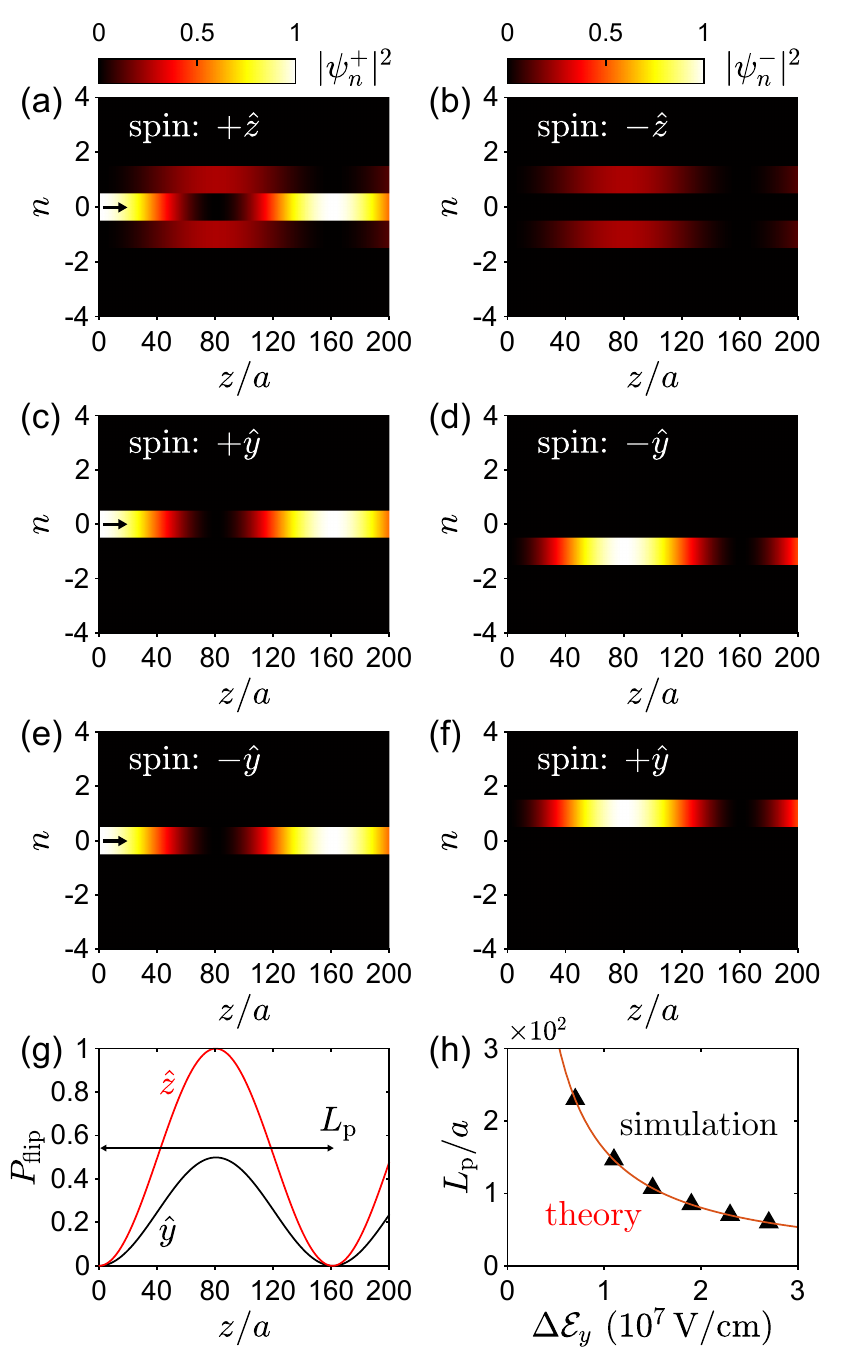}
		\par\end{centering}
	\caption{ (a)-(f) Distributions of the (a),(c),(e) spin-preserving ($|\psi_n^+|^2$) and (b),(d),(f) spin-flipped electrons ($|\psi_n^-|^2$) among discrete energy levels (plotted in the color scale as ribbons) for different propagation distances, calculated within the diffractionless approximation ($x=$const.). The input electron beams are spin-up polarized (see black arrows, i.e., $\psi_0^+$) along (a) $+\hat{z}$,  (c) $+\hat{y}$, and (e) $-\hat{y}$, respectively. (g) The spin-flip probabilities for incident electron spins along $\hat{x}$ and $\hat{y}$ show the same period of $L_{\rm p}$. (h) The period $L_{\rm p}$ for different $\Delta\mathcal{E}_y$ (see definition in Fig.\ \ref{Fig1}). We assume an electric field component $\Delta\mathcal{E}_y=1\times10^7\,$V/cm in (a)-(g), and use parameters $\lambda=1\,\mu$m and $a=100\,$nm for all calculations.
	}
	\label{Fig2}
\end{figure}

\begin{figure*}[t]
	\begin{centering}
		\includegraphics[width=1\textwidth]{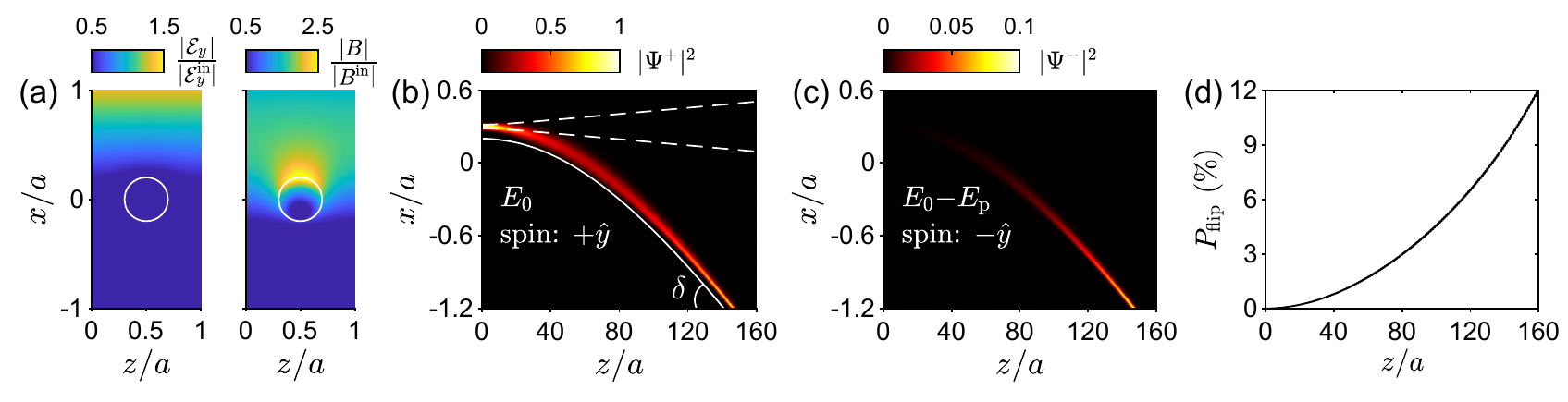}
		\par\end{centering}
	\caption{ (a) Electric (right) and magnetic (left) near field surrounding a nanowire array excited by a plane wave of wavelength $\lambda=1\,\mu$m. We assume a platinum nanowire array, with an experimentally measured optical permittivity $-105.76+16.49 i$ \cite{WGA09}, radius of $R=20\,$nm, and a lattice constant of $a=100\,$nm. (b)-(d) The spatial distributions of the (b) spin-preserving and (c) spin-flipped electron density, and the (d) spin-flip probability, calculated for an electron beam interacting with the optical near field shown in (a). An electron beam, spin-polarized along $+\hat{y}$ and with a phase-matched velocity, i.e., $\beta=1/10$, is initially focused at $z=0$ and $x=0.3a$, with a waist of $3\,$nm. In (c), the spin-flipped electrons are scattered to the state of energy $E_p-E_0$. The unperturbed beam width is shown by the white dashed curves in (b). The electron beam is slightly deflected due to a reshaping effect in the optical field, and the nanowire array is adaptively rearranged to avoid crossing (the top of the nanowires are shown by the white solid curve). The incident electric field is assumed to be $\mathcal{E}_y^{\rm in}=4\times 10^7\,$V/cm.  }
	\label{Fig3}
\end{figure*}

\section{Diffractionless approximation}
To concisely reveal the spin-flip physics of an electron beam in optical near fields, we first proceed with calculations based on the diffractionless approximation ($\hat{p}^2_x\approx 0$), assuming that the electron dynamics on different straight trajectories ($x=$const.) are independent. With this approximation, the electron spin dynamics is simply governed by a one-dimensional (1D) equation reduced from Eq.\ (\ref{Schr2}). The numerical results calculated by the reduced 1D equation are shown in Fig.\ \ref{Fig2}, where we adopt the simplified matrix elements in Eq.\ (\ref{Mmat}) and assume spin-polarized incident electron beams.

Since the magnetic near-field component  $\Delta \Bb$ is circularly polarized, the electron spin dynamics in the x-z plane is isotropic. Here, we chose the incident electron spin along $+\hat{z}$ to demonstrate the electron spin dynamics in the x-z plane (see Appendix), as shown in Fig.\ \ref{Fig2}(a) and (b). During the interaction with the near field, the occupancy of the initial electron state, $|\psi_0^+(z=0)|^2=1$, gradually decreases [Fig.\ \ref{Fig2}(a)], while the adjacent energy levels $\psi_{\pm 1}^\pm$ of both spin directions along $\pm\hat{z}$ [see Fig.\ \ref{Fig2}(a) and (b)] are equally populated via inelastic scattering. As the electron beam propagates further, the populations show an oscillatory behavior, while still retaining only these above-mentioned states, $\psi_0^+$ and $\psi_{\pm 1}^\pm$, populated. This contrasts with the PINEM interactions, as the electron populations in PINEM constantly spread out and randomly walk \cite{DGH21} among energy levels.

The incident electron spin along $+\hat{z}$ gives rise to a symmetric distribution of electrons among energy states, due to the lack of a symmetry-breaking mechanism. By contrast, for an incident electron spin along $\pm \hat{y}$, the electron spin is parallel to the spin AM of the near field, as revealed by the circular polarization of $\Delta \Bb$. In this configuration, the electron chirally couples to the field, leading to asymmetric patterns of the energy state populations [Fig.\ \ref{Fig2}(c)-(f)]. For the incident electron spin along $+\hat{y}$, the possible transition from the initial state is that the electron simultaneously releases energy quanta of $\hbar\omega$ and AM of $\hbar$ into the optical near field, namely $\psi_0^+\rightarrow \psi_{-1}^-$ [Fig.\ \ref{Fig2}(c) and (d)]. Likewise, the incident electron spin along $-\hat{y}$ only gives rise to the transition $\psi_0^+\rightarrow \psi_{+1}^-$ [Fig.\ \ref{Fig2}(e) and (f)]. Intriguingly, this type of asymmetric inelastic scattering offers a pathway for polarizing electron beams in a mixed state. More precisely, for an unpolarized incident electron beam, the electrons scattered to states of energy $E_{n+1}$ and $E_{n-1}$ are spin polarized along $+\hat{y}$ or $-\hat{y}$, respectively, and can be further separated by an energy filter. 

The energy level occupancy [Fig.\ \ref{Fig2}(a)-(f)] and the spin-flip probability [Fig.\ \ref{Fig2}(g)], defined by $P_{\rm flip}= |\Psi^-|^2/|\Psi|^2$, both exhibit oscillatory behaviors with the same period $L_{\rm p}$. At the distance of $L_{\rm p}/2$, where the initial state $\psi_0^+$ is fully depleted, for the incident electron spin along $+\hat{x}$ and $+\hat{y}$, the total electron population is equally allocated to states $\psi_{\pm 1}^\pm$ or totally transferred to the state $\psi_{-1}^-$, leading to a maximum $P_{\rm flip}$ of 0.5 and 1, respectively [see Fig.\ \ref{Fig2}(g)]. The repetition of the population among states observed here is, in fact, a magnetic Rabi oscillation, and the period can be estimated by $L_{\rm p}=2\pi v_0/\Omega_{\rm r} $, where the magnetic field induces a Rabi frequency of $\Omega_{\rm r}= 4\mu_B \Delta \mathcal{E}_y /\beta \hbar$ [Fig.\ \ref{Fig2}(h)].

\section{Spin-flip of diffractive electron beams}
To include the diffraction effects, we perform 2D calculations of Eq.\ (\ref{Schr2}), taking into account the transverse momentum (the $\hat{p}^2_x$ term) and the pondermotive force. In the calculation, we adopt a numerically solved electromagnetic field [see Fig.\ \ref{Fig3}(a)], where $\Bb$ is enhanced near the nanowire. We consider here a Gaussian electron beam, spin-polarized along $+\hat{y}$ and propagating though the near field calculated above. The electron beam is focused at $z=0$ with a distribution of $\Psi(x,z=0)=e^{-(x-30\,\text{[nm]})^2/w_0^2}$, where $w_0=3\,$nm is the beam waist. The unperturbed Gaussian beam width, $w(z)=w_0\sqrt{1+(z/z_R)^2}$, is indicated by the dashed curves in Fig.\ \ref{Fig3}(b), where $z_R=\pi w_0^2/\lambda_e$ and $\lambda_e=2\pi \hbar/ p_0 $ is the electron wavelength. 

When interaction with the electromagnetic field is considered, the Gaussian electron beam is reshaped due to an additional phase imprinted by the ponder-motive force (see, e.g., \cite{paper368}). According to Eq.\ (\ref{Schr2}), for the TE field considered in this work, the ponder-motive force introduces a phase, captured by
$d \varphi= e q^2|\mathcal{E}_y|^2 dz /2 m_e v_0$. In this regard, the electron wave segment propagating in a region of higher intensity accumulates a larger phase, so that the electron beam is deflected toward the nanowires surrounded by lower field intensity [see $|\mathcal{E}_y|$ in Fig.\ \ref{Fig3}(a)]. To avoid crossing of the electron beam with the nanowires, as shown by the solid curve of $x_j+R$ in Fig.\ \ref{Fig3}(b), we adaptively displace each nanowire slightly in the x-direction, while maintaining the distance between the center of each nanowire, $(x_j,z_j)$ ($j$th nanowire), and the electron-density peak in the plane $z=z_j$.  With this adaptive design, the electron beam experiences an acceleration along the x-direction, thereby following a parabolic-like trajectory. Additionally, the reshaping effect leads to a better confinement and even re-focusing of the electron beam. We note that the paraxial approximation assumed to derive Eq.\ (\ref{Schr2}) is valid, since the electron beam deflection angle, $\delta\approx 1.8a/120a\approx 15\,$mrad [see Fig.\ \ref{Fig3}(b)], is still small.

Despite the reshaping effect, our adaptive design of the nanowire array guarantees a strong interaction between the electron beam and the magnetic field localized around the nanowire [see $|\Bb|$ in Fig.\ \ref{Fig3}(a)], which gives rise to efficient spin-flip transitions. As shown in Fig.\ \ref{Fig3}(c) and (d), the population of the spin-flipped electrons with energy $E_0-E_p$ [see Fig.\ \ref{Fig2}(d)] continuously increases along the propagation, and the spin-flip probability $P_{\rm flip}$ reaches $\sim 12\%$ at a propagation distance of $16\,\mu$m. Similar spin-flip processes can also be observed for an incident electron spin in the x-z plane, but with the spin-flip probability $P_{\rm flip}$ merely halved, as can be anticipated from Fig.\ \ref{Fig2}(g). According to our discussions of Fig.\ \ref{Fig2}(c)-(f), for an unpolarized incident electron beam, when all other parameters are the same as in Fig.\ \ref{Fig3}, at the propagation distance of $16\,\mu$m, $12\%$ of the electrons in the beam are scattered to the two states of energies $E_0+ E_p$ and $E_0- E_p$ (see Appendix) ($6\%$ in each state). The scattered electrons in these two states have opposite spins in the y-direction and can be further separated by an electron spectrometer.

\section{Concluding remarks}
In this work, we propose a spin-flip effect of electrons in optical near fields. This effect is exceptionally efficient compared to previous proposals in free space (see Appendix). In addition to requiring an ultra-strong laser intensity, the previous schemes that use free-space light to flip electron spins or polarize electron beams also face other experimental difficulties, such as measurement stability and electron--laser overlap under extreme conditions \cite{PTS18,CBG18}. The spin-flip effect proposed in the present work is feasible, since PINEM is already a well-established technology that has even been recently implemented using a scanning electron microscope \cite{SCH22}. The incident laser intensity adopted in Fig.\ \ref{Fig3}, $I^{\rm in}= 8.5\times 10^{12}\,$W/cm${^2}$, is attainable using commercial ultrafast lasers, and the intensity $I^{\rm in}$ required to achieve the same $P_{\rm flip}$ as in Fig.\ \ref{Fig3} can be further reduced by extending the interaction length. As the typical laser damage threshold of metals is several J/cm$^2$ and the plasmon-enhanced absorption is absent for the TE illumination, the nanostructures should sustain a laser illumination with a pulse duration of $\sim 100$s$\,$fs, provided the peak intensity of $I^{\rm in}= 8.5\times 10^{12}\,$W/cm${^2}$. In addition, in future experiments, the use of a higher input electric current can further reduce the laser intensity required to induce observable spin-flipped electrons and prove the spin-flip effect proposed in this study. 

\appendix

\section{Derivation of Eq. (1) in the main text}
In the main text, Eq. (1) is used to describe the spin-flip and beam-diffraction effects. We provide a detailed derivation of the more general form of Eq. (1), which is capable of describing paraxial electron dynamics in an arbitrary electromagnetic field. As explained in the main text, in the non-relativistic limit, the evolution of an electron wave function involving the spin degrees of freedom can be fully captured with Pauli's equation,
\begin{align}
	(\frac{\pb^2}{2 m_e}+\mathcal{\hat{H}}_{\rm I})\Psi(\rb,t)=i\hbar\partial_t\Psi(\rb,t) , \label{Pauli}
\end{align} 
where the interaction Hamiltonian is explicitly written as
\begin{align}
	\mathcal{\hat{H}}_{\rm I}=&\frac{e}{2 m_e c}\left[\Ab(\rb,t)\cdot\pb + \pb \cdot \Ab(\rb,t) \right] \nonumber\\
	&+ \mu_B \Bb(\rb,t)\cdot\vec{\sigma} +\frac{e^2}{2m_e c^2} \Ab^2(\rb,t) . \nonumber
\end{align} 
The Pauli equation can be derived from the Dirac equation in the non-relativistic limit. This approximation can also introduce the spin-orbit interaction term, which can be neglected in our work, as we investigate only the paraxial electron beams here. In fact, in most studies on electron-spin separation, the spin-orbit interaction effect always provides a minor contribution. In Eq.\ (\ref{Pauli}), we also choose the radiation gauge, so that the scalar potential vanishes, and the electric field is related to the vector potential by $\vec{\mathcal{E}}(\rb,t)=-\partial_t \Ab(\rb,t)/c$. In this work, we also deal only with monochromatic electromagnetic fields, and for all the field quantities, we use the following notation, e.g., $\vec{\mathcal{E}}(\rb,t)=\vec{\mathcal{E}}(\rb) e^{-i\omega t}+c.c.$

Within the paraxial approximation, the electron dynamics on different straight lines is weakly coupled; therefore, we can expand the electron wave function on such straight lines using the 1D plane wave eignstates,
\begin{align}
	\Psi (\rb,t)=	\sum_{s=\pm} \Psi^s (\rb,t) \ket{s}= \sum_{n,s=\pm} \psi_n^s (\rb) e^{\frac{i}{\hbar}(p_n z-E_n t)} \ket{s},  \nonumber
\end{align} 
where $p_n=\sqrt{2m_e E_n}$ is the canonical momentum of the electron in the state of energy $E_n$, and $\ket{s}$ is the spinor denoting the electron spin. Inserting this expansion into the Pauli equation [Eq.\ (\ref{Pauli})] and eliminating those terms corresponding to the non-perturbative Schrodinger equation, we find
\begin{widetext}
\begin{align}
	\sum_{ns}  e^{\frac{i}{\hbar}(p_{n} z-E_{n} t)} \bigg[ \frac{\hat{p}_x^2}{2 m_e}   + \tilde{v}_{n} \hat{p}_z
	+ \frac{e}{c} v_{n} A_z(\rb,t)   + \mu_B \Bb(\rb,t)\cdot\vec{\sigma} +\frac{e^2}{2m_e c^2} \Ab^2(\rb,t)    \bigg] \psi_{n}^{s} (\rb)\ket{s}  =0  ,\label{expand}
\end{align} 
where $v_n=p_n/m_e$ is the canonical velocity of the electron in the state of energy $E_n$, and $\tilde{v}_n=v_n + e  A_z(\rb,t)/m_e c$ is the corresponding kinetic velocity. To derive the equation above, we also neglect the terms containing $\hat{p}_z^2$ and $\hat{p}_z A_z(\rb,t)$, assuming that the electron wave amplitude $\psi_n(\rb)$ and the electromagnetic field amplitude $A_z(\rb)$ vary slowly in space with respect to the phase term $e^{ip_n z}$, 

We then multiply Eq.\ (\ref{expand}) by the eigenfunction $e^{-i(p_{n''} z-E_{n''} t)/\hbar}\bra{s''}$ and perform a summation over the introduced index $(n'',s'')$. By retaining only the time-independent terms, Eq.\ (\ref{expand}) is rewritten as
\begin{align}
	0=&	\sum_{n''s''} \Bigg\{ \sum_{n s }  \delta_{nn''}^{ss''}\left[ \frac{\hat{p}_x^2}{2 m_e} + \tilde{v}_n \hat{p}_z  +\frac{e^2}{m_e c^2} |\Ab(\rb)|^2  \right]  \psi_n^s (\rb)   \nonumber\\
	&+\sum_{n's'}\delta_{n',n''-1} e^{\frac{i}{\hbar}(p_{n'} -p_{n''} )z} \left[ \delta_{s's''} \frac{e}{c} v_{n'}  A_z(\rb)   
	+  \bra{s''}\mu_B \Bb(\rb)\cdot\vec{\sigma}\ket{s'}  \right]      \psi_{n'}^{s'} (\rb)  \nonumber\\
	&+ \sum_{n's'}\delta_{n',n''+1}e^{\frac{i}{\hbar}(p_{n'} -p_{n''} )z} \left[ \delta_{s's''}\frac{e}{c} v_{n'}  A_z^*(\rb)    
	+ \bra{s''}\mu_B \Bb^*(\rb)\cdot\vec{\sigma}\ket{s'}  \right]   \psi_{n'}^{s'} (\rb) \Bigg\}  . \nonumber
\end{align}
All other time-oscillating terms are neglected, because their contributions are averaged out along the propagation of electrons. The equation above includes a set of coupled differential equations, which can be rewritten in the matrix form,
\begin{align}
	\big( \tilde{v}_n \hat{p}_z  +\frac{\hat{p}^2_x}{2 m_e}\big) \psi_n^s(\rb)= - \sum_{n's'} \mathcal{M}^{ss'}_{nn'} \psi_{n'}^{s'}(\rb) ,\label{matrix}
\end{align} 
where the nonzero matrix elements of $\mathcal{M}^{ss'}_{nn'}$ are
\begin{align}
	\mathcal{M}_{n, n-1}^{ss'}&= e^{\frac{i}{\hbar} (p_{n-1}-p_n) z} \left[ -\delta_{ss'} \frac{ie}{\omega}v_{n-1} \mathcal{E}_z(\rb)+ \mu_B \bra{s} \Bb(\rb) \cdot\vec{\sigma}\ket{s'} \right]  , \nonumber \\ 
	\mathcal{M}_{n, n+1}^{ss'}&= e^{\frac{i}{\hbar} (p_{n+1}-p_n) z} \left[ \delta_{ss'} \frac{ie}{\omega}v_{n+1} \mathcal{E}_z^* (\rb)+ \mu_B \bra{s} \Bb^*(\rb) \cdot\vec{\sigma}\ket{s'} \right]  , \nonumber \\ 
	\mathcal{M}_{n n}^{ss}& = \frac{e^2}{ m_c \omega^2} |\Eb(\rb)|^2      .\nonumber
\end{align} 
\end{widetext} 
In the previous PINEM theory, the field contribution in the canonical momentum is regarded as small, and thus, $\tilde{v}_n$ is approximated by $v_n$. More safely, for the transverse electric field examined in our work, the component of the vector potential parallel to the electric field is absent; thus, $\tilde{v}_n= v_n$ is rigorously satisfied. In the case of nonrecoil limit, i.e., $E_0\gg \hbar\omega$, the equation above can be further simplified using the following approximations, $v_n \approx v_0$ and $p_{n+1}-p_{n}\approx  \hbar \omega/v_0$. Equation (\ref{matrix}) finally reduces to Eq.\ (1) in the main text. In addition, the matrix $\mathcal{M}$ in the equation above does not seem to be Hermitian, because we neglect the term $\hat{p}_z A_z(\rb,t)$ in Eq.\ (\ref{expand}). Nevertheless, the Hermicity is recovered in the nonrecoil approximation.

\section{Comparison of spin-flip efficiency with previous works}
In Fig.\ 3, we chose a laser intensity of $8.5\times 10^{12}\,$W/cm${^2}$, while in previous studies that used free-space lasers to induce the electron spin flips, the adopted intensities were mainly within the range of $10^{18}$--$10^{22}\,$W/cm$^2$ (see the references in the second paragraph of the main text). Incidentally, a few works even adopted lower laser intensities than our work. The laser intensities in these studies varied by several orders, because the assumed interaction lengths $L$ and the achieved spin-flip probabilities were different. 

To make a reasonable comparison of the spin-flip efficiencies in different works, we define the efficiency by averaging the spin-flip probability $P_{\rm fl}$ over the interaction length $L$ and laser intensity $I^{\rm in}$, i.e.,
\begin{align}
	\eta=\frac{P_{\rm fl}}{I^{\rm in} L^2}, \label{eff}
\end{align}
where $I^{\rm in}$ is the incident laser intensity. In this definition, the spin-flip probability is averaged over $I^{\rm in} L^2$, because the transition amplitude should be proportional to $|\mathcal{E}^{\rm in} L|$ in the weak interaction regime, according to Eq.\ (\ref{matrix}). Although the results in Fig.\ (3) are beyond the weak interaction regime, we can still try to apply Eq. (\ref{eff}) to Fig.\ 3 and find that the corresponding spin-flip efficiency is $\eta\approx 5.5 \times 10^{-9}\,$W$^{-1}$.

As the next step, we chose a work \cite{MHS15} that used the lowest incident laser intensity as an example to demonstrate that our results greatly improve the spin-flip efficiency. In \cite{MHS15}, regarding the spin-flip effect in the regular Kapitza-Dirac (KD) effect using two lasers with the same frequency, also referred to as the spin-Kapitza-Dirac (SKD) effect, the spin-flip efficiency was found to be very low. More precisely, for an electron velocity of $1\times 10^7$m/s and an interaction length of $1\,$mm, an illumination intensity of approximately $10^{14}\,$W/cm$^2$ finally results in a spin-flip probability of $1.3\times 10^{-3}$. As a comparison, if their work chose the same interaction length of $16\,\mu$m and incident laser intensity of $8.5\times 10^{12}\,$W/cm$^2$ as in Fig.\ 3, the spin-flip probability should be further reduced to 
\begin{align}
	P_{\rm fl}&\approx 1.3\times 10^{-3} \times \left(\frac{16[\mu \text{m}]}{1[ \text{mm}]}\right)^2 \times \frac{8.5\times 10^{12}[\text{W/cm}^2]}{10^{14}[\text{W/cm}^2]} \nonumber\\
	   &= 2.8 \times 10^{-8}\ll 12\%.  \nonumber
\end{align}
The comparison above can also be made equivalently using the spin-flip probability we defined in Eq.\ (\ref{eff}), which leads to a spin-flip efficiency of $\eta \approx 1.3\times 10^{-15}\,\text{W}^{-1}$ for the SKD effect, relative to $\eta\approx 5.5 \times 10^{-9}\,$W$^{-1}$ in our study.

In addition, in \cite{MHS15}, a scheme based on the two-color KD effect was also proposed to realize the electron spin flip, where the spin-flip efficiency is much higher than that in the SKD. According to \cite{MHS15}, for an electron velocity of $1\times 10^7$m/s and an interaction length of $1\,$mm, an illumination intensity of approximately $10^{11}\,$W/cm$^2$ results in a spin-flip probability of $7.4\times 10^{-4}$. Compared with our results, the spin-flip efficiency is found to be $\eta \approx 7.4\times 10^{-13}\,\text{W}^{-1}$, which is still smaller by several orders than our work.

\section{Three-dimensional electron beam}
In the main text, we consider a 2D Gaussian electron beam, which consists of the Fourier components of different wave vectors (in the x-z plane) perpendicular to the transverse electric field $\mathcal{E}_y$. For such a 2D electron beam, spin-preserving PINEM scattering is absent, which can be caused only by the electric field parallel to the electron wave vector [see Eq.\ (\ref{matrix})]. However, in a realistic experiment, the incident electron wave function should be three-dimensionally (3D) confined, with a Gaussian beam profile also in the y-z plane, as illustrated in Fig.\ \ref{FigS1}. Unlike the 2D situation, the 3D electron beam also includes Fourier components of the wave vectors that are not perpendicular to $\mathcal{E}_y$. As illustrated in Fig.\ \ref{FigS1}, for a Fourier plane wave of wave vector $\kb_e$, the electric field $\mathcal{E}_y$ includes a component parallel to $\kb_e$, which can then give rise to PINEM scatterings. For simplification, in our discussions below we focus on the Fourier components of the wave vectors in the y-z plane; i.e., $k_e^x=0$. In the following, we compare the strengths of the spin-flip and PINEM interactions and reveal the critical beam waist when spin-flip scattering is dominant.

\subsection{Strength comparison: Spin-flip vs. PINEM interactions}
To compare the interaction strengths of spin-flip and PINEM scattering, we first consider a 2D electron beam along $+\hat{z}$ through an optical near field that includes electric field components of $\mathcal{E}_y$ and $\mathcal{E}_z$, as in the main text. We also focus on the spin flip along the y-direction, so the term $\mu_B \Bb \cdot \vec{\sigma}$ results only in the spin-flip scattering. In this configuration, according to the corresponding matrix elements in Eq.\ (\ref{matrix}), the strength of the PINEM interaction induced by $\mathcal{E}_z$ is characterized by $\chi_{\rm PINEM}=|ev_0\mathcal{E}_z/\omega|$. Similarly, for a large wave vector in the near field ($g\gg q$), the x- and z-components of the magnetic field contribute equally; thus, so the interaction strength of the spin-flip transition along the y-direction can be evaluated with $\chi_{\rm flip}=|2\mu_B g  \mathcal{E}_y|$, according to Eq.\ (3) in the main text. 

\begin{figure}
	\begin{centering}
		\includegraphics[width=0.5\textwidth]{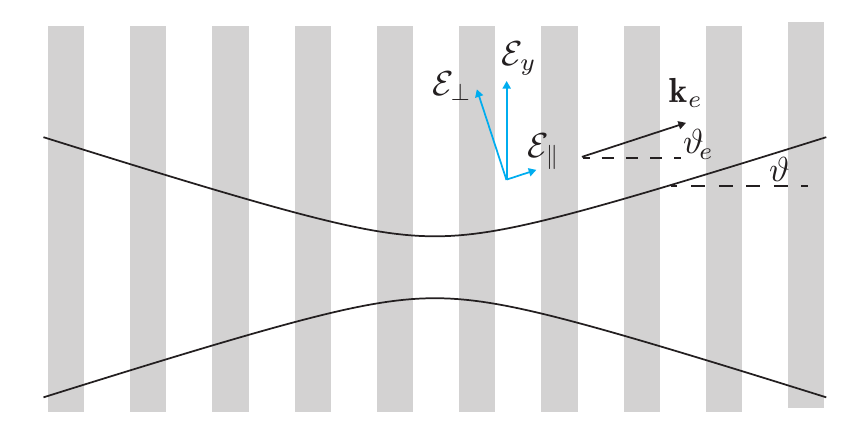}
		\par\end{centering}
	\caption{Illustration of a three-dimensional Gaussian electron beam propagating through the optical near field of a nanowire array, as in Fig.\ 1 in the main text.}
	\label{FigS1}
\end{figure}

Similarly, for the interaction scenario between a near field that includes only $\mathcal{E}_y$ and a Fourier component of a wave vector $\kb_e=(k_e^x,k_e^z)$ (see Fig.\ \ref{FigS1}), the interaction strengths can be found equivalently. In this situation, the spin-flip and PINEM interaction strengths can be found by simply substituting $\mathcal{E}_y$ and $\mathcal{E}_z$ in $\chi_{\rm flip}$ and $\chi_\mathcal{\rm PINEM}$, using $\mathcal{E}_\perp$ and $\mathcal{E}_\parallel$, i.e., the components perpendicular and parallel to $\kb_e$, respectively. Eventually, we can find the ratio
\begin{align}
	\frac{\chi_{\rm flip}}{\chi_\mathcal{\rm PINEM}}= \frac{g}{k_0} \frac{\mathcal{E}_\perp}{\mathcal{E}_\parallel} =\frac{\hbar\omega }{m_e c^2} \left( \frac{v_0}{c}\right)^{-2} \frac{\cos(\vartheta_e)}{\sin(\vartheta_e)}. \label{eta}
\end{align} 
This equation again demonstrates that a slow electron is preferable for observing the proposed spin-flip effect, as explained in the main text. Intuitively, the factor $g/k_0$ in the equation above reveals that the spin-flip transition becomes more prominent as the photonic scale approaches that of the electron. 

\subsection{Critical Gaussian beam waist along the y-direction}
For a transverse electric field $\mathcal{E}_y$ ($\mathcal{E}_z=0$), we can find a certain critical angle $\vartheta_{\rm cr}$, at which the interaction strengths of the spin-flip and PINEM scattering are equal; i.e., $\chi_{\rm flip}/\chi_{\rm PINEM}=1$. For the parameters adopted in Fig.\ 3, the corresponding critical angle is easily found to be
\begin{align}
	\vartheta_{\rm cr} \approx   0.24 \, [\text{mrad}]. \nonumber
\end{align}
Alternatively, for a 3D incident Gaussian electron beam, when measured at a far field angle $<\vartheta_{\rm cr}$, we should observe a stronger spin-flip effect than the spin-preserved PINEM effect. 

We can also define a critical waist of the Gaussian beam in the y-direction, at which the electron beam is dominated by spin-flip scattering. As shown in Fig.\ \ref{FigS1}, the lateral diffraction angle of the beam in the y-z plane is $\vartheta=\lambda/\pi w_0^y$, where $w_0^y$ is the beam waist along the y-direction, and $\lambda_e=0.24\,[\text{\AA}]$ is the electron wave length corresponding to the velociy $v_0=c/10$. When the diffraction angle of the electron beam coincides with the critical angle, $\vartheta =\vartheta_{\rm cr}$, spin-flip scattering significantly dominates the electron beam compared to the PINEM interactions. For the parameters in Fig. 3, the corresponding critical waist is found to be $w_0^{y,{\rm cr}}=\lambda_e/\pi \vartheta_{\rm cr}=32\,[\text{nm}]$. By further increasing the waist $w_0^{y}$, the electron beam is more similar to a plane wave, and the corresponding spin dynamics should be more consistent with the results in the main text.

\newpage
\begin{widetext}
\section{Supplemental calculations}
\begin{figure}[h]
	\begin{centering}
		\includegraphics[width=0.6\textwidth]{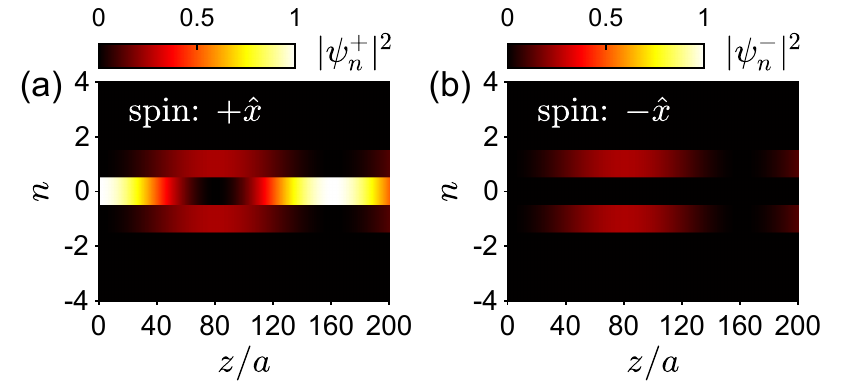}
		\par\end{centering}
	\caption{Same as Fig.\ 2(a) and (b) for the incident electron spin along $+\hat{x}$.}
	\label{FigS2}
\end{figure}

\begin{figure}[h]
	\begin{centering}
		\includegraphics[width=0.95\textwidth]{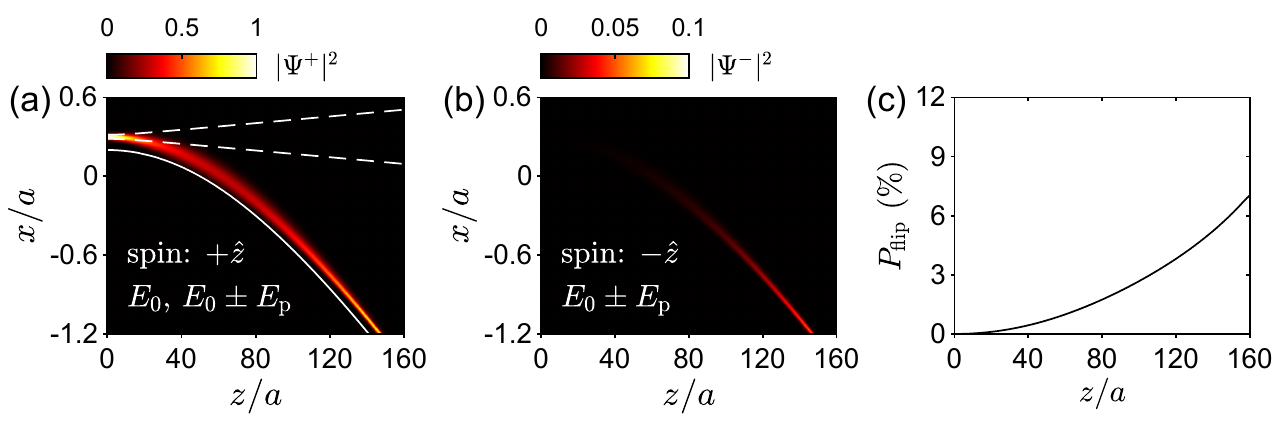}
		\par\end{centering}
	\caption{Same as Fig.\ 2(b)--(d) for the incident electron spin along $+\hat{z}$. Different from Fig.\ 2(b), the spin-preserving electrons (a) are distributed on the states of energy $E_0$ and $E_0\pm E_p$ [see Fig.\ 2(a)]. Similarly, the spin-preserving electrons (b) are distributed on the energy levels, $E_0\pm E_p$ [see Fig.\ 2(b)]. }
	\label{FigS2}
\end{figure}
\end{widetext}

\end{document}